\documentclass{revtex4}

\newcommand{\lag}{{\mathcal L}}
\newcommand{\lsim}{\lower.7ex\hbox{$\;\stackrel{\textstyle<}{\sim}\;$}}

\usepackage[dvips]{graphicx}

\begin{document}

\begin{flushright}
hep-ph/0111065\\
UCD-2001-11
\end{flushright}

% You should use BibTeX and revtex.bst for references
\bibliographystyle{revtex}

% Use the \preprint command to place your local institutional report
% number  and your conference paper identification number on the
% title page in preprint mode. Multiple \preprint commands are allowed.
%\preprint{}

%Title of paper
\title{Complementarity of muon-conversion and linear collider-based
experiments for lepton-flavor violating $U(1)$ gauge bosons}
% Optional argument for running titles on pages
%\title[]{}

% repeat the \author .. \affiliation  etc. as needed
% \email, \thanks, \homepage, \altaffiliation all apply to the current
% author. Explanatory text should go in the []'s, actual e-mail
% address or url should go in the {}'s for \email and \homepage.
% Please use the appropriate macro for the type of information

% \affiliation command applies to all authors since the last
% \affiliation command. The \affiliation command should follow the
% other information

\author{Brandon Murakami and James D.~Wells}
\email[]{murakami@physics.ucdavis.edu, jwells@physics.ucdavis.edu}
%\homepage[]{Your web page}
%\thanks{}
%\altaffiliation{}
\affiliation{{\it University of California, Davis, California 95616}}

%Collaboration name if desired (requires use of superscriptaddress
%option in \documentclass). \noaffiliation is required (may also be
%used with the \author command).
%\collaboration{}
%\noaffiliation

\date{\today}

\begin{abstract}
% insert abstract here
In general, attempts to extend the Standard Model will include extra
gauge structure.  We parameterize string and technicolor models for a
$Z'$ boson with primitive lepton flavor violating interactions.
Calculations for its muon conversion rate ($\mu^-N \to e^-N$) on
Titanium are made and used to show the potential of forthcoming
experiments MECO at Brookhaven National Laboratory and PRIME at the
Japan Hadron Facility (to be renamed).  For reasonable choices of
parameters, such
$U(1)$ bosons with masses the order 10 TeV and $\mu eZ'$ charge as low as
${\sim}10^{-5}$ are demonstrated to be
accessible for MECO and PRIME.  Also, a demonstration of the
complementarity of parameter space coverage for future colliders and
muon conversion experiments is given.
\end{abstract}
% insert suggested PACS numbers in braces on next line
% \pacs{}

%\maketitle must follow title, authors, abstract and \pacs
\maketitle

% body of paper here - Use proper section commands
% References should be done using the \cite, \ref, and \label commands
%\section{}
%\label{}
%\subsection{}
%\subsubsection{}

In the grander endeavors to unify the forces, explain the spectrum of
fermion masses, and understand the hierarchy problem, additional gauge
and vector bosons generally arise from string models, GUTs, $N\geq2$
supersymmetry, technicolor models,
and models with extra dimensions.  Models with extended gauge groups,
such as string models \cite{Nardi:1993nq, Cleaver:2000xe,
Chaudhuri:1995cd} and technicolor \cite{Rador:1999is, Muller:1996dj,
Yue:2000fh},
have potential to produce additional $U(1)'$ (or $Z'$) gauge bosons
with flavor-dependent couplings to the fermions.  Lepton flavor
violation (LFV) is allowed in many extensions of the standard 
model~\cite{Feng:2000ci}.  LFV may occur at tree-level
bosonic exchange by either flavor off-diagonal $Z'$ charges or
indirectly through the standard model $Z$ boson mixing with the
$Z'$. The additional fermions that arise from extended gauge models
may further promote LFV currents by mixing with the standard model
(SM) leptons \cite{Bernabeu:1993ta}.  Such a boson's mass may be on
the order of 1 TeV in models such as those of the aforementioned
references, prompting attention at collider experiments.

Muon conversion experiments refer to those that create a ground state
muonic atom which is allowed to then decay.  Much like, say, cryogenic
WIMP detection or muon $g-2$ experiments, muon conversion experiments
have great potential to unveil new physics.  Currently SINDRUM II at
the Paul Scherrer Institut has attained the current lowest exclusion
limit of  $B \equiv \Gamma(\mu^-N \to e^-N)/\Gamma(\mu^-N \to \nu_\mu
N') \leq 6.1 \times 10^{-13}$ \cite{Wintz}.  They are currently taking
data that should improve their limit by an order of magnitude.  The
forthcoming MECO (Muon Electron COnversion) experiment (E940 at BNL)
has plans for construction in early 2002 and data acquisition in 2006
\cite{Molzon}.  They will have potential to probe down to $5\times10^{-17}$.
The Japan Hadron Facility began construction in the spring of 2001 and
requires 6 years to complete \cite{Kuno}.  It will include a
high-intensity muon source, PRISM, which will feed the PRISM Mu-E
conversion (PRIME) experiment.   At this time of writing, PRIME is in
the proposal stage.  They have plans to reach a sensitivity of
$10^{-18}$, more than four orders of magnitude improvement over
the present limit.

Other models may contribute entirely or partially to muon
conversion.  Through loop effects of squarks and sleptons,
supersymmetry is well known to produce LFV interactions
\cite{Martin:1997ns}.  Loop effects with neutrino mixing would not be
sufficient to explain a muon conversion signal at MECO or PRIME
\cite{Petcov:1977ff}.  However, in a supersymmetric model
\cite{Hisano:1996cp, Borzumati:1986qx} or model with singlet neutrinos
propagating in extra dimensions \cite{Faraggi:1999bm}, neutrino masses
may be held accountable.  Models with a +2 electrically charged Higgs
boson also exhibit muon conversion potential through loop diagrams
with charged leptons provided a $y_{ij} \phi^{++} \bar{e}_{Ri}^c e_{Rj}$
interaction exists \cite{Raidal:1998hq}.  Other exotic scenarios are
reviewed in Ref. \cite{Kuno:1999jp}.  In our calculations, we assume
muon conversion is due entirely to LFV $Z'$ bosons. 

Any model with a $Z'$ that has flavor couplings not proportional to
the identity matrix in which the fermions are in their ``natural
basis'' (before Yukawa diagonalization) will have a non-zero $\mu e
Z'$ vertex in the physical basis.  Let us call this a flavor-dependent
vector-boson $G_\mu$.  These models will
provide the coupling constant $g_G$ and charges that form six
matrices $q^{\psi}_{ij}$ for the fermions $\psi$---the left- and
right-handed lepton, up-type quarks, and down-type quarks.  The
$Z-G$ mixing angle $\theta$ and the mass $m_G$ are free
parameters.  In our analysis, other parameters are the six unitary matrices
$U^{\psi}$ that perform the Yukawa rotations with the constraint
$V_{\rm CKM} \equiv U^{u_L\dagger}U^{d_L}$.  The charges in the
physical basis are denoted $Q^{\psi}_{ij} \equiv
U^{\psi\dagger}_{ik}q^{\psi}_{kl}U^{\psi}_{lj}$.  We express the
$G$-fermion interactions as
\begin{equation}
\lag \supset \frac{g_G}{\sin\theta_{\rm W}} \sum_{\psi'}
\bar\psi'_i \gamma^\mu (q^{\psi'_L}_{ij} P_L + q^{\psi'_R}_{ij} P_R)
\psi'_j G_\mu
\end{equation}
where $\psi'$ denotes the fermions in the natural basis, and
\begin{equation}
\lag \supset \frac{g_G}{\sin\theta_{\rm W}} \sum_\psi \bar\psi_i
\gamma^\mu (Q^{\psi_L}_{ij} P_L + Q^{\psi_R}_{ij} P_R) \psi_j G_\mu
\end{equation}
where the fermions are in the physical basis.

Technicolor bosons may have similar parameterization.
Ref. \cite{Lynch:2001md} reviews such models that replace a $U(1)_Y$
or $SU(2)_L$ of the standard model with a pair of gauge groups
$U(1)_{\rm l} \times U(1)_{\rm h}$ or $SU(2)_{\rm l} \times SU(2)_{\rm
h}$.  The gauge group with the ``l'' (``light'') subscript is assigned
SM-like couplings to the first two fermion families, and the other
(``h'' for ``heavy'') receives SM-like couplings to the third
generation.  The $U(1)_{\rm l} \times U(1)_{\rm h}$ symmetry is broken
by a new Higgs boson at some scale higher than the weak scale to form
a $U(1)_Y$ representation.  In either case, after properly enforcing
electroweak symmetry breaking there will be an additional $U(1)'$
boson with generation-diagonal couplings such that the first two
generations have equal $U(1)'$ charge but different than the
third. The relative strength of these charges are determined
completely by the mixing angle of the $Z$ and $Z'$.  If these gauge
groups are considered to act on the fermions in the natural basis,
there will, in general, be generation off-diagonal LFV $G$ couplings
in the physical basis.

With the objective of choosing the minimum number of parameters while
maintaining arbitrary LFV couplings and boson mass, we define our model
as follows.  Additional fermions are necessary to cancel the anomalies
that arise with an additional gauge boson; their effects are ignored.
For the purposes of illustrating the $G$ boson
effects on LFV interactions in the most conservative way, 
the $Z-G$ mixing angle $\theta$ is considered negligible, and the
mixing of the kinetic gauge terms, i.e.\ $\frac{\chi}{2} Z^{\mu\nu}
G_{\mu\nu}$, is also omitted.  Upper limits on phenomenological
constraints on $|\chi|^2$ are on the orders of $10^{-6}$ for gauge
mediated SUSY breaking and $10^{-16}$ for gravity mediated SUSY
breaking \cite{Dienes:1997zr}.  Relative to the amplitude for
tree-level $G$-exchange in $t$-channel muon conversion, the amplitude for kinetic
mixing would include $|\chi|^2$  suppression, and so is generally
negligible anyway.

In further efforts to be conservative on the effects of generation-dependent
$Z'$'s, we set the first two generations of right-handed lepton charges equal
and the charge matrix diagonal,
$q^{l_R} = {\rm diag}(q^{l_R}_{11},\, q^{l_R}_{11},\, q^{l_R}_{33})$
(in the natural basis).  These assignments will allow the
$\mu e G$ charge $Q^{l_R}_{12}$ to be non-zero and have an arbitrary value.
We set the left-handed lepton charges to zero
$q^{l_L}_{ij} = 0$.  Other gauge interaction choices (i.e.\ purely
axial vector, vector and axial vector, mixed helicities, etc.) would
have no impact on our study. We choose the $G$-quark interactions to
be $Q_{11}^{u_L}=1$, $Q_{11}^{u_R}=1$, $Q_{11}^{d_L}=1$, and
$Q_{11}^{d_R}=1$.  All charges not noted have no direct relevance for our
calculations.

For the Yukawa unitary rotation matrices, we choose as many possible
to be trivial, $U^{l_L}=U^{u_R}=U^{d_L}=U^{d_R}=1$.  $U^{u_L} = V_{\rm
CKM}$ is necessary to meet the definition of the CKM matrix.  The
meaningful rotation is assigned to $U^{l_R}$ here, and 
has the parameterized form of the CKM matrix,
\begin{equation}
U^{l_R} =
\left( \begin{array}{ccc}
c_{12}c_{13} & s_{12}c_{13} & s_{13} \\
-s_{12}c_{23}-c_{12}s_{23}s_{13} & c_{12}c_{23}-s_{12}s_{23}s_{13}
& s_{23}c_{13} \\
s_{12}s_{23}-c_{12}c_{23}s_{13} & c_{12}s_{23}-s_{12}c_{23}s_{13}
& c_{23}c_{13}
\end{array} \right).
\end{equation}
The notation $s_{ij}$ and $c_{ij}$ means sines and cosines of
parameters $\theta_{12}$, $\theta_{23}$, $\theta_{13}$ which need not
be the CKM values.  We have ignored the allowed complex phase for
simplicity.  

Explicitely, the right-handed leptons have charges
\begin{equation}
{Q^{l_R}} = U^{l_R\dagger} q^{l_R} U^{l_R}.
\end{equation}
Note, if $U^{l_R}\simeq V_{\rm CKM}$ we would expect an estimated
value of $|Q^{l_R}_{12}|\lsim 10^{-4}$.  If it is expected that the
lepton generation-transitions should follow in a similar manner as the
neutrino mixing, the large mixing angle (LMA) solution and ``low
probability, low mass'' (LOW) solution both yield roughly
$|Q^{l_R}_{12}|\lsim 0.1$ \cite{Fukugita:2001rk}.

The muon conversion rate is \cite{Langacker:2000ju}
\begin{eqnarray}
B &=& \frac{G_F^2\alpha^3 m_\mu^5}{2\pi^2\Gamma_{\rm capture}}
\frac{Z_{\rm eff}^4}{Z}|F_P|^2
\left( |Q_{12}^{l_L}|^2 +|Q_{12}^{l_R}|^2 \right) \nonumber\\
&& \times
\left| \frac{g_G}{g_Y}\sin\theta\cos\theta
\left(1-\frac{m_W^2}{m_G^2\cos^2\theta_W}\right) \left[
\frac{1}{2}(Z-N)-2Z\sin^2\theta_W \right] \right. \nonumber\\
&& \left. + \frac{g_G^2}{g_Y^2} \left( \sin^2\theta +
\frac{m_W^2}{m_G^2\cos^2\theta_W} \cos^2\theta \right) \left[
(2Z+N)(|Q_{11}^{u_L}|^2 + |Q_{11}^{u_R}|^2) \right.\right. \nonumber\\
&&+ \left.\left. (Z+2N)(|Q_{11}^{d_L}|^2 + |Q_{11}^{d_R}|^2)
\right] \right|^2,
\end{eqnarray}
and for $^{48}$Ti, nuclear form factor $F_P=0.54$
\cite{Bernabeu:1993ta}, $Z_{\rm eff}=17.6$ \cite{Sens}, and the muon
capture rate $\Gamma_{\rm capture}=2.6 \times 10^{-6}$ s$^{-1}$
\cite{Suzuki:1987jf}.  For this model's parameterization, the muon
conversion rate is 
\begin{equation}
B = 0.157 \left(\frac{g_G}{g_Y}\right)^4 |Q^{l_R}_{12}|^2
\left( \frac{1\,{\rm TeV}}{m_G} \right)^4.
\end{equation}

Fig.\ 1 shows the potential of muon conversion experiments.  The
reach for extremely heavy bosons or moderately massive bosons with
extremely small LFV couplings is intriguing.  For example, a 10 TeV
$G$ boson with off-diagonal couplings $\frac{g_G}{g_Y} Q^{l_R}_{12}$ as
small $10^{-5}$ may be implied by a signal at MECO or PRIME.
Furthermore these experiments have potential to imply $G$ bosons of
astonishing high masses:  roughly from $\cal O$(10 TeV) to $\cal
O$(1,000 TeV) for reasonable choices of off-diagonal couplings.  With
the promise of strict bounds or detection, this exemplifies the
importance of muon conversion experiments.

\begin{figure}[!btp]
\center
\resizebox{6 in}{!}{\includegraphics{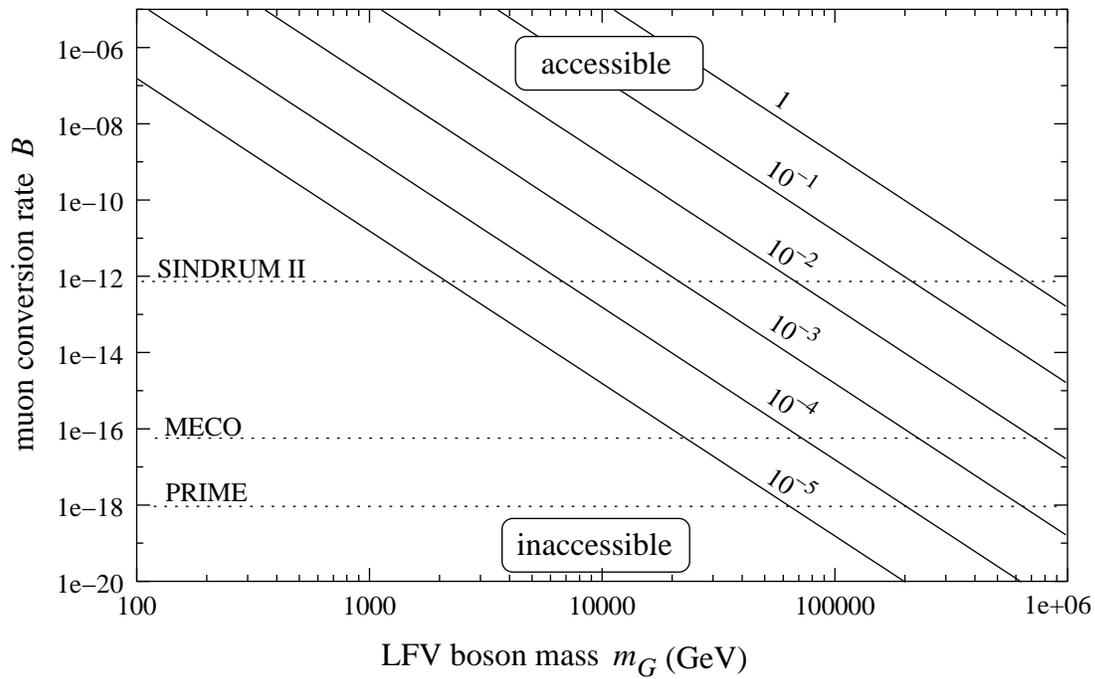}}
\caption{LFV boson mass $m_G$ (GeV) vs.\ the muon conversion rate $B
\equiv \Gamma(\mu^-N \to e^-N)/\Gamma(\mu^-N \to \nu_\mu N')$.  The
potential for muon conversion experiments to access our model's $G$
boson is demonstrated here.  The solid diagonal lines represent
different values of the $\mu e G$ vertex couplings $Q^{l_R}_{12}$.
The horizontal dotted lines represent the limits set
or to be set by muon conversion experiments.  The information here is
best appreciated by noting the suprisingly large boson mass and
small couplings possibly accessible by experiment.}
\end{figure}

Fig.\ 2 demonstrates how a high-energy linear collider (LC) 
would complement muon conversion experiments.  Together, both programs can
cover all but the ``quadrant'' of small coupling $Q^{l_R}_{12}$ and
large mass $m_G$.  With precision measurements of $e^+e^- \to
\mu^+\mu^-$, measuring a 1\% difference in the standard model cross
section sets limits on what masses of a $G$ boson may be implied.  The
contributions from the $t$-channel exchange (which utilize the LFV
vertex) are relevant only for the parameter
space region of relatively small off-diagonal charge $Q^{l_R}_{12}$
and low boson mass, which is ruled out by SINDRUM II.  Naively, one
might expect $e^+e^- \to e\mu$ to be a better probe, however the
limits from this process would also be ruled out by SINDRUM II.  For
small charge $Q^{l_R}_{12}$, the mass limits are
approximately independent of the coupling.  The Tevatron should be able
to continue its search for $Z'$ gauge bosons above 1 TeV, and the
LHC will be able to discover most $Z'$ bosons with mass above
about 5 TeV. Combining the $\mu -e$ LFV searches with the collider searches
provides a much enhanced ability to search for new gauge symmetries.

For the $G$ boson model, the rate for the magnetic interaction $\mu
\to e\gamma$ is zero by
virtue of all left-handed lepton charges being zero and the chirality
flips inherent in magnetic moments.  Therefore, $\mu \to
e\gamma$ does not constrain this model.  Nevertheless, if we
chose a model with non-zero left- and right-handed lepton charges,
such as a model with purely vectorial couplings, muon conversion
experiments would generally still be more useful 
in constraining lepton-violating
$Z'$ models.

The current $\mu^+ \to e^+e^+e^-$ limit is $\Gamma(\mu\to
eee)/\Gamma(\mu\to e\nu\bar\nu) = 1.0\times10^{-12}$, held by SINDRUM
\cite{Bellgardt:1988du}.  This limit is competitive with the current
SINDRUM II limit for muon conversion.  For example, a 10 TeV $G$ boson
would be at the verge of being seen in $\mu\to eee$ if it had charges
of $Q^{l_R}_{11} = 1$ and $Q^{l_R}_{12} = 5.4\times10^{-3}$.
Next generation $\mu -e$ conversion experiments would then
far outpace current constraints from $\mu\to eee$.  We extend this
study in Ref.~\cite{Murakami:2001cs} to a more complete study,
including discussions of $\mu\to e\gamma$, $\mu\to eee$, the muon 
anomalous magnetic moment, and other relevant lepton observables.

\begin{figure}[!btp]
\center
\resizebox{6 in}{!}{\includegraphics{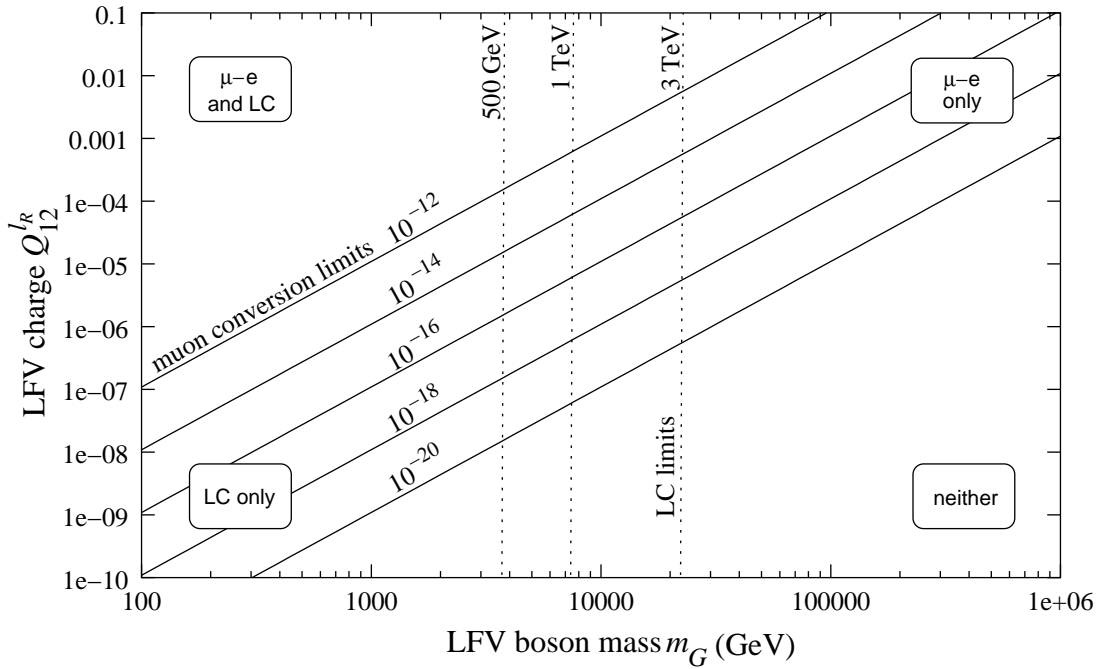}}
\caption{LFV boson mass $m_G$ (GeV) vs. the $\mu eG$ vertex charge
$Q^{l_R}_{12}$.  The vertical lines show the limit of a 1\% difference
in $\Delta\sigma(e^+e^- \to \mu^+\mu^-) / \sigma_{SM}(e^+e^- \to
\mu^+\mu^-)$ where $\Delta\sigma$ is the difference of the cross section
of our extended model and standard model. The diagonal lines show the
charge $Q^{l_R}_{12}$ necessary for a muon conversion experiment to
detect a $G$ boson of a certain mass provided the experiment has the
ability to reach the muon conversion rate noted on the plot.  These
plots show linear collider and muon conversion experiments to be
highly complementary:  1)  In the upper left quadrant, both muon
conversion experiments and an NLC will detect our model's boson.  2)
In the upper right quadrant, only MECO and PRIME have potential.  3)
In the lower left quadrant, only an NLC has potential. 4) In the lower
right quadrant, neither have potential.}
\end{figure}

%%%%%%%%%%%%%%%%%%%%%%%%%%%%%%%%%%%%%%%%%%%%%%%%%%%%%%%%%%%%%%%%%%%%%%%

% figures should be put into the text as floats.
% Use the graphicx package (distributed with LaTeX2e).
% See the LaTeX Graphics Companion by Michel Goosens, Sebastian Rahtz,
% and Frank Mittelbach for instance.
%
% Here is an example of the general form of a figure:
% Fill in the caption in the braces of the \caption{} command. Put the label
% that you will use with \ref{} command in the braces of the \label{} command.
%
% \begin{figure}
% \includegraphics{}%
% \caption{}
% \label{}
% \end{figure}

% tables follow here or maybe be put in the text
%
% Here is an example of the general form of a table:
% Fill in the caption in the braces of the \caption{} command. Put the label
% that you will use with \ref{} command in the braces of the \label{} command.
% Insert the column specifiers (l, r, c, d, etc.) in the empty braces of the
% \begin{tabular}{} command.
%
% \begin{table}
% \caption{}
% \label{}
% \begin{tabular}{}
% \end{tabular}
% \end{table}

% If you have acknowledgments, this puts in the proper section head.
\begin{acknowledgments}
% put your acknowledgments here.
We thank K. Dienes and S. Mrenna for helpful
discussions.  This work was supported by the Department of Energy and
the Alfred P. Sloan Foundation.
\end{acknowledgments}

% Create the reference section using BibTeX:
\bibliography{your bib file}

\end{document}